\documentclass[amsmath,amssymb,aps,superscriptaddress]{revtex4-1}
\usepackage[margin=1in]{geometry}			
\usepackage{graphicx,epsfig}
\usepackage{xcolor}
\usepackage{hyperref}
\hypersetup{
    colorlinks = true,
}

\begin{document}
\author{M.-G.~Hu}
\thanks{These two authors contributed equally.}
\affiliation{Department of Chemistry and Chemical Biology, Harvard University, Cambridge, Massachusetts, 02138, USA.}
\affiliation{Department of Physics, Harvard University, Cambridge, Massachusetts, 02138, USA.}
\affiliation{Harvard-MIT Center for Ultracold Atoms, Cambridge, Massachusetts, 02138, USA.}

\author{Y.~Liu}
\thanks{These two authors contributed equally.}
\affiliation{Department of Physics, Harvard University, Cambridge, Massachusetts, 02138, USA.}
\affiliation{Department of Chemistry and Chemical Biology, Harvard University, Cambridge, Massachusetts, 02138, USA.}
\affiliation{Harvard-MIT Center for Ultracold Atoms, Cambridge, Massachusetts, 02138, USA.}

\author{D.~D.~Grimes}
\affiliation{Department of Chemistry and Chemical Biology, Harvard University, Cambridge, Massachusetts, 02138, USA.}
\affiliation{Department of Physics, Harvard University, Cambridge, Massachusetts, 02138, USA.}
\affiliation{Harvard-MIT Center for Ultracold Atoms, Cambridge, Massachusetts, 02138, USA.}

\author{Y.-W.~Lin}
\affiliation{Department of Chemistry and Chemical Biology, Harvard University, Cambridge, Massachusetts, 02138, USA.}
\affiliation{Department of Physics, Harvard University, Cambridge, Massachusetts, 02138, USA.}
\affiliation{Harvard-MIT Center for Ultracold Atoms, Cambridge, Massachusetts, 02138, USA.}

\author{A.~H.~Gheorghe}
\affiliation{Department of Physics, Harvard University, Cambridge, Massachusetts, 02138, USA.}

\author{R. Vexiau}
\affiliation{Laboratoire Aim\'{e} Cotton, CNRS, Universit\'{e} Paris-Sud, ENS Paris-Saclay, Universit\'{e} Paris-Saclay, 9145 Orsay cedex, France}

\author{N. Bouloufa-Maafa}
\affiliation{Laboratoire Aim\'{e} Cotton, CNRS, Universit\'{e} Paris-Sud, ENS Paris-Saclay, Universit\'{e} Paris-Saclay, 9145 Orsay cedex, France}

\author{O. Dulieu}
\affiliation{Laboratoire Aim\'{e} Cotton, CNRS, Universit\'{e} Paris-Sud, ENS Paris-Saclay, Universit\'{e} Paris-Saclay, 9145 Orsay cedex, France}

\author{T.~Rosenband}
\affiliation{Department of Physics, Harvard University, Cambridge, Massachusetts, 02138, USA.}

\author{K.-K.~Ni}
\email[To whom correspondence should be addressed. E-mail: ]{ni@chemistry.harvard.edu}
\affiliation{Department of Chemistry and Chemical Biology, Harvard University, Cambridge, Massachusetts, 02138, USA.}
\affiliation{Department of Physics, Harvard University, Cambridge, Massachusetts, 02138, USA.}
\affiliation{Harvard-MIT Center for Ultracold Atoms, Cambridge, Massachusetts, 02138, USA.}

\title{{\Large Direct observation of bimolecular reactions of ultracold KRb molecules}}
\date{\today}
\begin{abstract}

\begin{large}

Femtochemistry techniques have been instrumental in accessing the short time scales necessary to probe transient intermediates in chemical reactions. Here we take the contrasting approach of prolonging the lifetime of an intermediate by preparing reactant molecules in their lowest ro-vibronic quantum state at ultralow temperatures, thereby drastically reducing the number of exit channels accessible upon their mutual collision. Using ionization spectroscopy and velocity-map imaging of a trapped gas of potassium-rubidium molecules at a temperature of 500~nK, we directly observe reactants, intermediates, and products of the reaction $^{40}$K$^{87}$Rb + $^{40}$K$^{87}$Rb $\rightarrow$ K$_2$Rb$^*_2$ $\rightarrow$ K$_2$ + Rb$_2$. Beyond observation of a long-lived energy-rich intermediate complex, this technique opens the door to further studies of quantum-state resolved reaction dynamics in the ultracold regime.

\end{large}

\end{abstract}
\maketitle
\large

The creation of ensembles of molecules at ultralow temperatures enables a variety of high resolution spectroscopic studies, allows broader exploration of reaction phase space,  and promises quantum state control over the outcome of chemical reactions. Already, investigations of single partial wave collisions have provided detailed benchmarks of short-range molecular potentials~\cite{henson2012observation,yang2019observation}, exotic conditions at low temperatures have facilitated the synthesis of new chemical species~\cite{puri2017synthesis}, and highly sensitive and precise methods of detection have traced state-to-state reactions between atoms and weakly-bound Feshbach molecules~\cite{hoffmann2018reaction}. Further, chemical reaction rates for barrierless reactions can be altered~\cite{ospelkaus2010quantum,kilaj2018observation}, in some case by orders of magnitude, merely by changing the nuclear spins of the reactants and entering quantum degeneracy~\cite{de2019degenerate}. These studies all rely on the extraordinary control attainable over the quantum states of the ultracold molecules.

\begin{figure}
\centering
\includegraphics[width=4 in]{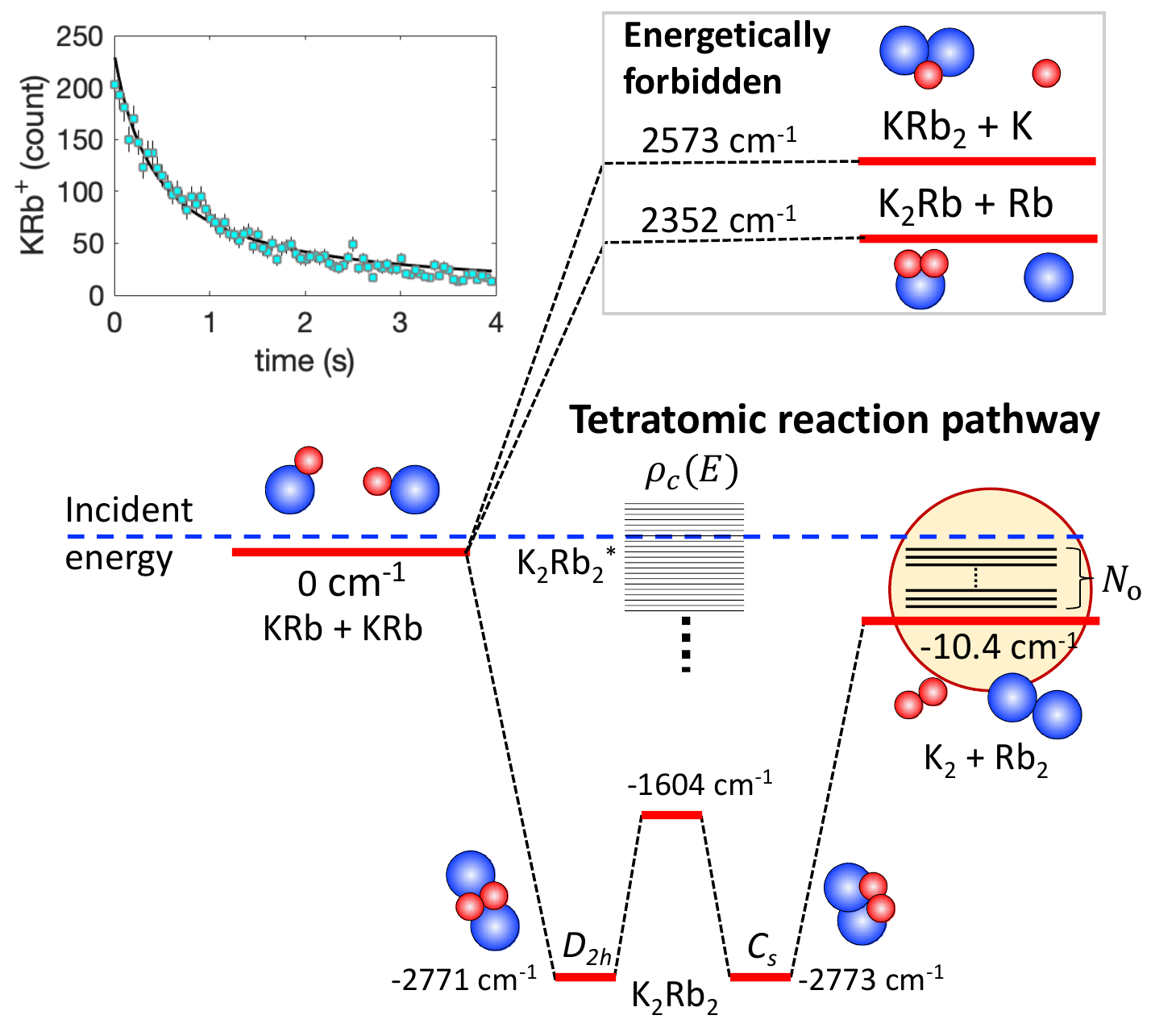}
        \caption{\textbf{Energetics of the bimolecular reactions of ultracold KRb molecules.} The ground-state energies are obtained from spectroscopic data for KRb \cite{ni2008high}, K$_2$ \cite{falke2006sigma} and Rb$_2$ \cite{amiot1990laser} and from calculation for KRb$_2$, K$_2$Rb, and K$_2$Rb$_2$ at the equilibrium configuration \cite{byrd2010structure}. Here we define the incident energy of two free KRb molecules as zero energy. Because the energies of the triatomic reaction channels are much higher than that of the reactants, these channels are energetically forbidden. In comparison, the tetratomic reaction channel KRb + KRb $\rightarrow$ K$_2$Rb$^*_2$ $\rightarrow$  K$_2$ + Rb$_2$ is exothermic and therefore energetically allowed. K$_2$Rb$^*_2$ denotes the transient intermediate complex. $\rho_c(E)$ is the density of states of K$_2$Rb$^*_2$ near the incident energy $E$. Two isomers of K$_2$Rb$_2$ with $D_{2h}$ and $C_s$ symmetries connect to the KRb+KRb and K$_2$+Rb$_2$ dissociation limits, respectively. $N_o$ is the number of exit channels that consist of all combinations of quantum states of K$_2$ and Rb$_2$ that have a total energy below $E$. The inset in the top-left shows the number decay of KRb molecules measured using ionization detection. Each data point is accumulated over $300$ experimental cycles. The error bars denote shot noise. The black curve is a weighted fit to the two-body decay model used in \cite{ospelkaus2010quantum} with the root mean squared error (RMSE) being $1.37$.}
\label{fig1}
\end{figure}

Despite recent advances in ultracold molecule studies, a key capability has been missing: namely, characterization of transient reaction intermediates and final products. Previous experiments have shown evidence of ultracold reactions between bialkali molecules through the quantum-state-specific detection of loss of reactants~\cite{ospelkaus2010quantum}, similar to that shown in the inset to Fig.~\ref{fig1}, giving insights into how long-range forces determine the kinetic collision rates of the reactants. These reactions have been observed to occur with a high probability after just a single collision, approaching unity in certain cases~\cite{ospelkaus2010quantum,ye2018collisions,gregory2019are}. Despite tour-de-force levels of control over the precise ro-vibrational quantum state of the reactants to open up additional energetically allowed reaction channels, no significant differences based on the reactant species or initial quantum state have yet been observed~\cite{ye2018collisions,gregory2019are}, and the nature of the molecular loss is still under debate~\cite{christianen2019trapping}.

When two molecules approach one another, they initially form an energy-rich intermediate collision complex, the dynamics of which could hold the key for understanding the details of ensuing these ultracold, barrierless, bimolecular reactions. In higher-temperature reactions, this transient complex only persists for one or two vibrational periods, and at most on the order of a rotational period ($\sim1$ ps)~\cite{bauer1979four,miller1972molecular}. Studying the dynamics or kinetics of such complexes in the gas phase has typically required ultrafast~\cite{brooks1988spectroscopy,zewail2000femtochemistry,gruebele1991femtosecond,sims1992femtosecond,polanyi1995direct} or stabilizing collisional~\cite{womack2015observation,bjork2016direct,bui2018direct} techniques. Structural investigations of these complexes have been previously obtained by photodetachment~\cite{garand2008nonadiabatic, continetti2017dynamics}, photoabsorption~\cite{su2013infrared}, or photodissociation~\cite{sato2001photodissociation}. 
Based on the Rice-Ramsperger-Kassel-Marcus (RRKM) theory, the lifetime of an intermediate complex is given by $\tau_c=2\pi\rho_c/N_o$, where $\rho_c(E)$ denotes the density of states of the intermediate complex near the incident energy, $E$, and $N_o$ is the number of energetically allowed exit channels (Fig.~\ref{fig1}). Preparing reactant KRb molecules in the pure ro-vibronic ground state in the ultralow temperature regime tightly constrains the number of energetically allowed exit channels, greatly extending the lifetime of the intermediate complex. For reactions between bialkali molecules, depending on the species, $\tau_c$ has been estimated to be on the order of hundreds of nanoseconds to microseconds~\cite{mayle2013scattering,christianen2019quasiclassical}, which makes direct observation of the complex a possible goal. However, no such observations have been made because all previous work has been based on the observation of loss of reactants. Direct multi-species detection methods are necessary in order to fully describe the details of these ultracold reactions~\cite{nesbitt2012toward}.

Here, we report the direct detection of a predicted intermediate as well as products in the ultracold chemical metathesis reaction $^{40}$K$^{87}$Rb + $^{40}$K$^{87}$Rb $\rightarrow$ K$_2$Rb$_2^*$  $\rightarrow$ K$_2$ + Rb$_2$ (see Fig.~\ref{fig1})~\cite{mayle2013scattering, christianen2019quasiclassical}. We combine precise quantum-state preparation of the ultracold reactants with an ionization-based detection method that allows for direct and simultaneous detection of reactants (KRb), intermediates (K$_2$Rb$_2^*$), and final products (K$_2$, Rb$_2$).

\begin{figure}
\centering
\includegraphics[width=6.5 in]{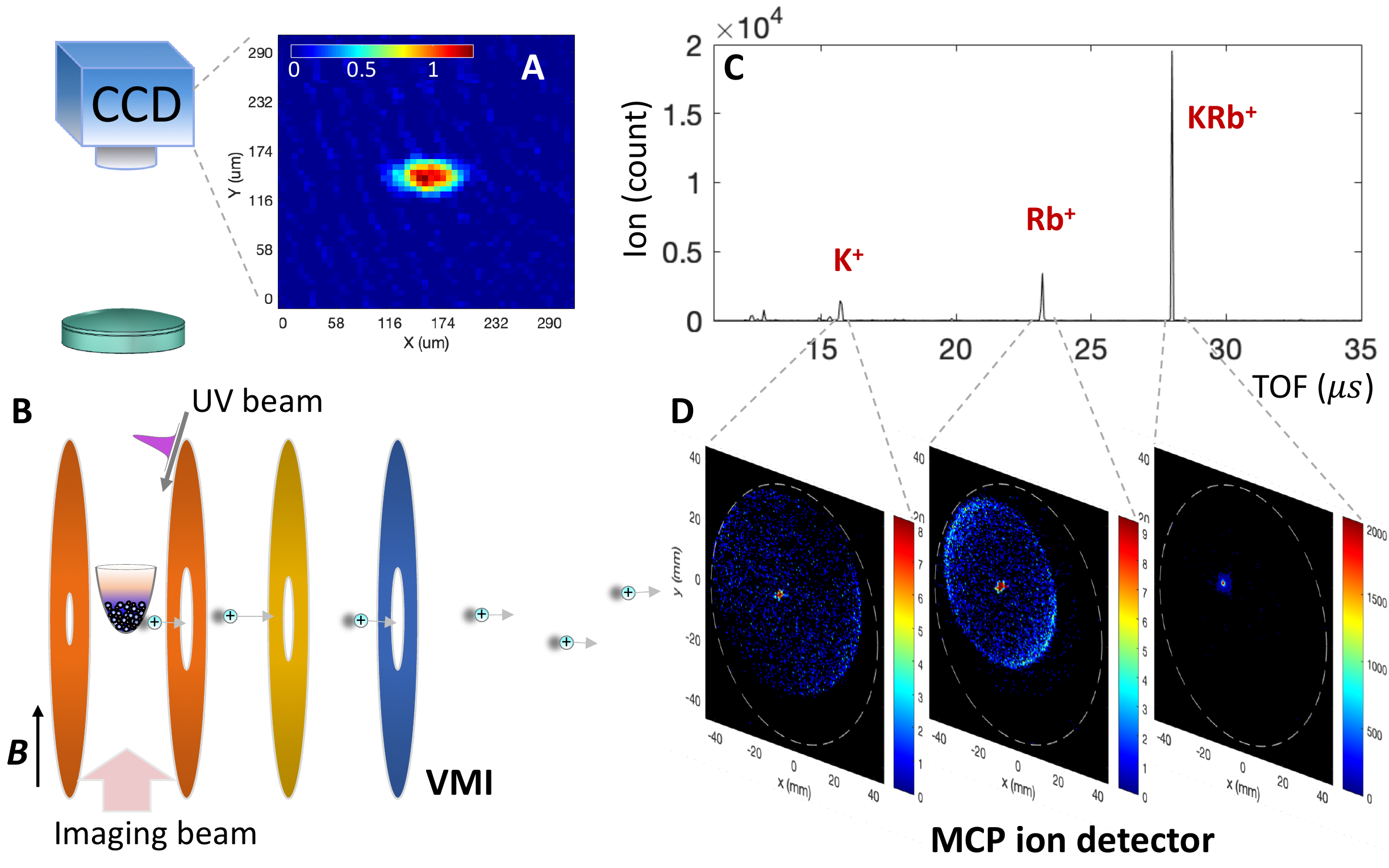}
    \caption{\textbf{Schematic of our ultracold chemistry apparatus.} Ground-state KRb molecules at $500$ nK are trapped by a crossed optical dipole trap. (\textbf{A}) An absorption image of KRb molecules. The colorbar indicates the optical depth of the KRb cloud. (\textbf{B}) These trapped molecules are surrounded by velocity-map imaging (VMI) ion optics~\cite{eppink1997velocity}, which consist of a series of disk-shaped electrodes. We use a pulsed UV laser to photoionize the molecules. (\textbf{C}) An example TOF spectrum, which can be converted to a mass spectrum using the relation, $\text{mass} = 0.16248 (\text{amu}/\mu\text{s}^2) \times \text{TOF}^2$. (\textbf{D}) For each species identified in the mass spectrum we also obtain a VM image, from which the momentum distribution can be inferred.}
\label{fig2}
\end{figure}

We began by implementing an established protocol~\cite{ni2008high} to create an optically trapped gas of $v=0,\,N=0$, $X^1\Sigma^+$ ground-state KRb molecules. Here $v$ and $N$ are the  vibrational and rotational quantum number of the molecules, respectively. In brief, ultracold K and Rb atoms are first converted to weakly-bound molecules with 20\% efficiency by a magnetic field sweep (1.4 ms) through a Feshbach resonance at 546.62~G~\cite{cumby2013feshbach}. Then a pair of Raman beams are applied in a stimulated Raman adiabatic passage (STIRAP) \cite{vitanov2001coherent} pulse sequence (4 $\mu$s) to coherently transfer the weakly-bound molecules into a single hyperfine state of the ro-vibronic ground state with 85\% efficiency. 
8~$\mu$s after the STIRAP pulse, we remove residual Rb and K atoms. Because the atom-to-molecule transfer is mostly coherent, we can reverse the transfer with high efficiency. To detect the ground-state KRb molecules, a reversed STIRAP sequence is applied followed by absorption imaging on an atomic transition (see example image shown in Fig.~\ref{fig2}A). Typically, $5\times 10^3$ KRb molecules are created at 500~nK with a peak density of $10^{12}$ cm$^{-3}$ and are trapped by a crossed optical dipole trap (ODT) at a laser wavelength of 1064~nm.

Because the  absorption imaging detection is tied directly to the quantum-state-specific STIRAP transfer, it is only sensitive to the  KRb molecules in the STIRAP populated quantum state. To probe chemical reaction products and the intermediate complex, we chose a more general detection method that entailed photoionization of neutral reaction species into bound molecular ions, acceleration of the ions in an electric field, and measurement of their arrival time and position on a multi-channel plate~(MCP) (Fig.~\ref{fig2}C). By combining mass spectrometry and velocity-map imaging (VMI)~\cite{eppink1997velocity} in our ultracold molecule apparatus, we could thereby identify reacting species and study reaction dynamics. 

We performed three separate experiments to probe the reactants, intermediate complex, and products of the ultracold reaction. The detection procedure worked as follows: After KRb creation but before the ionization pulse, we ramped the magnetic field down to 30~G within 15~ms to reduce subsequent Lorentz forces that might deflect ions away from the detector, housed 1~m downstream. We then applied an ultraviolet (UV) ionization pulse while simultaneously triggering the MCP to record ion signals. For the detection of reactants and products, we chose a photoionization wavelength of 285 nm, which is above the ionization threshold of KRb, K, Rb, as well as any species that consist of  combinations of multiple K and Rb atoms (table S1). For the detection of the intermediate complex, the wavelength was varied over the 285 to 356 nm range. To avoid space-charge effects, the laser power was kept low enough to ensure at most one ion was generated per UV pulse. The ODT was switched off for a variable time period during and before the ionization pulse to preclude its influence on the chemical reaction, the lifetime of the intermediate collisional complex, and the photoionization process.  We repeated this detection procedure at 1 kHz for the reactant and product detection (see timing diagram in fig.~S1), and at 7 kHz for the intermediate complex detection (see inset of Fig.~\ref{fig4}D). The mass, and thereby elemental composition, of each detected ion could be inferred from its time-of-flight (TOF), whereas the momentum of the ion was mapped through its location on the VM image \cite{SM}.

\begin{figure}
\centering
\includegraphics[width=5 in]{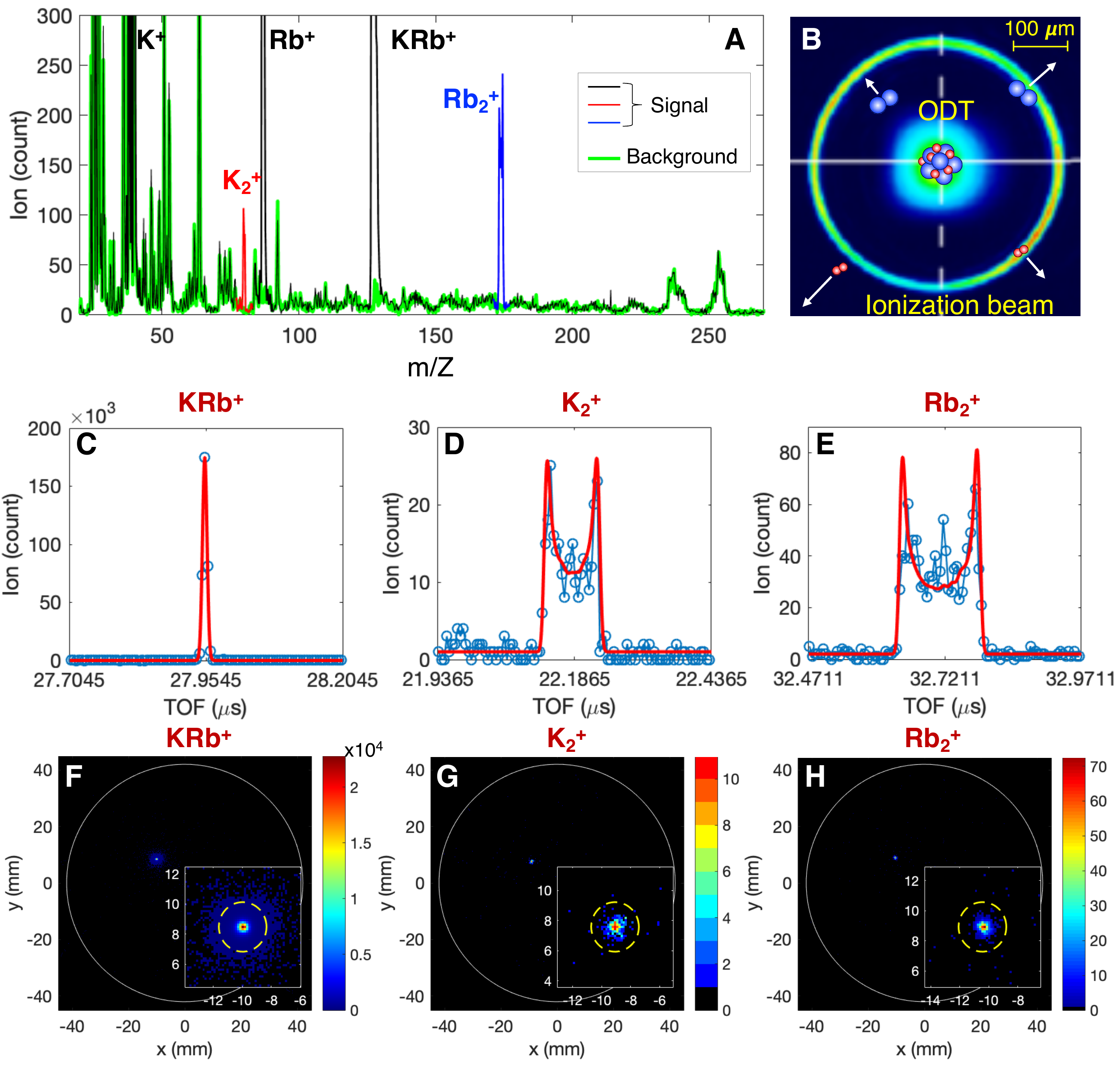}
    \caption{\textbf{Identification of the reaction products.} (\textbf{A}) The mass spectrum of the reaction products ionized by 285 nm UV laser pulses. The color-coded ion signals correspond to species associated with the reaction of two KRb molecules compared to the ionization background (the green trace). Noise ions that show up in both the signal and the background spectra have no significant effect on the ion signals of interest (Sec.~S6 in SM). (\textbf{B}) Geometries of the relevant beams with schematic representations of the reactants and products superimposed. The Gaussian beam spot in the center is the ODT and the ring surrounding it is the ionization beam. (\textbf{C-E}) The TOF data for KRb$^+$, K$_2^+$, and Rb$_2^+$ ions, respectively. The red curve in (\textbf{C}) is a time resolution limited Gaussian to describe TOF lineshape for the ions generated in the center, while those in (\textbf{D}) and (\textbf{E}) are simulated TOF lineshapes for the ions generated in the ring. For the simulation, we use physical parameters of our system such as the diameter of the hollow-bottle beams, 0.45 mm, the intersection angle of the two hollow beams, $40^{\circ}$, and the VMI electric field (sec. S3 in SM). The only fitting parameter in this model is the overall amplitude of the signal. (\textbf{F-H}) The momentum distributions of the KRb$^+$, K$_2^+$, and Rb$_2^+$ ions, respectively. The white solid circles represent the active area of the detector. The yellow dashed circles represent the momenta corresponding to $10.4\text{ cm}^{-1}$ of translational energy.}
\label{fig3}
\end{figure}

To demonstrate the ionization detection capability in our ultracold molecular apparatus and to gain information beyond absorption imaging, we first probed the trapped KRb molecules in the ODT (Fig.~\ref{fig2}C,D). As expected, the dominant signal results from KRb$^+$. The VM image for the KRb$^+$ signal has a width limited by the detector resolution, consistent with the negligible translational energy in the ultracold regime. Measurable amounts of Rb$^+$ and K$^+$ were also detected. The VM images for K$^+$ and Rb$^+$ both show two distinct components: an isotropic central peak and an anisotropic ring. The ions forming the central peak originate from residual ultracold atoms from the molecule creation process after the cleanup pulses. Based on the known ionization cross sections and estimated ion detection efficiencies (table S1), we put an upper bound of 250 atoms of each species in the trap. These populations are small compared to KRb, ensuring that the dominant reaction in the subsequent study is the desired bimolecular reaction. The sensitivity of ionization detection allowed us to quantify the small number of residual atoms in the ODT, which are not seen using absorption imaging. To analyze the Rb$^+$ ions forming the ring pattern, we extracted the translational energy release (TER) from the diameter of the ring to obtain a TER of $8.3\times10^{3}$ cm$^{-1}$. By comparing this TER to the calculated molecular potentials of KRb and KRb$^+$~\cite{korek2003theoretical} we identified a two-photon dissociative ionization pathway that contributes to this atomic ion signal. The same analysis also applies to the ring pattern of the K$^+$ ions (fig.~S3).

After KRb molecules are created, the bimolecular reaction occurs continuously with a measured decay rate coefficient of $7.6(3)\times10^{-12}$ cm$^3$/s until the reactants are depleted (Fig.~1 inset), consistent with previous studies~\cite{ospelkaus2010quantum}. To probe the products of the bimolecular reaction while reducing the perturbation to the reactants during ionization, we shaped our ionization beam into a ``hollow-bottle'' (Fig.~\ref{fig3}B) with the laser intensity concentrated in a ring outside of the ODT to keep the reactants in the dark; the measured intensity contrast between the peak and center of the beam was 500 \cite{SM}.  To further reduce the hollow volume for higher efficiency ionization, we crossed two hollow-bottle beams at a 40$^\circ$ angle centered on the ODT \cite{SM}. To observe the bimolecular reaction without the possible influence of the ODT light, we shut off the ODT for 170 $\mu$s prior to each ionization pulse, thereby precluding any role of the ODT in the formation of all but those products with translational energy less than 0.0127 $\text{cm}^{-1}$ \cite{footnote1}. 

The dominant peaks in the mass spectrum (Fig.~\ref{fig3}A) are again K$^+$, Rb$^+$, and KRb$^+$, primarily from photoionization of trapped KRb molecules by the residual intensity at the centers of the hollow-bottle beams. Aside from these dominant peaks, we can clearly identify ions corresponding to the masses of K$_2^+$ and Rb$_2^+$. All peaks aside from these five species appear with comparable intensities in a background spectrum (green trace) taken in the absence of ultracold atoms and molecules.

We postulate that K$_2^+$ and Rb$_2^+$ come from direct ionization of reaction products, K$_2$ and Rb$_2$ (Fig.~1).    To support such an assignment, we draw evidence from the TOF lineshapes and the VM images. The TOF lineshapes characterize the spatial origin of the ions in the ionization beam. The KRb$^+$ lineshape (Fig.~\ref{fig3}C) is sharp and described well by the time resolution limited Gaussian for ions that come from the central part of the hollow ionization beams, which coincides with the position of the ODT. K$_2^+$ and Rb$_2^+$ share similar TOF lineshapes, as shown in Figs. \ref{fig3}D \& E, which are much wider than that of KRb$^+$. The simulated lineshape (with only total amplitude as a free parameter,  see Sec. S3 in SM) based on the beam geometry for particles ionized by the ring portion of the hollow ionization beams matches well to the data, which supports the assignment that these signals is from reaction products escaping the central KRb cloud. The presence of a center peak in Fig.~\ref{fig3}E that is not captured by the simulated curve is likely due to the product ionization at the center of the hollow beams, where the beams are not perfectly dark. We also rule out the role of ion-neutral reactions due to their negligible estimated rates (Sec. S4 in SM).

In addition to the mass spectrometry of the K$_2^+$ and Rb$_2^+$ ions, we simultaneously recorded the momentum distribution of the K$_2^+$ and Rb$_2^+$ ions with VMI (Figs.~\ref{fig3}G \& H). To characterize the radius of the distribution, we performed Bayesian fits (Sec. S5 in SM) to the images, assuming a circular Gaussian density on a flat background with uninformative priors. The radius of K$_2^+$ (Rb$_2^+$) corresponds to a translational energy of 0.59 cm$^{-1}$ (0.29 cm$^{-1}$), well above the MCP resolution of 0.02 cm$^{-1}$. The ionization process of K$_2$ (Rb$_2$) would impart to the resulting ion a photon recoil energy of 0.0159 cm$^{-1}$ (0.0112 cm$^{-1}$), too small to significantly impact the momentum distribution of the ions. Therefore, the measured K$_2^+$ and Rb$_2^+$ translational energies closely resemble that of their parent neutrals.  The sum of measured translational energies is smaller than the exothermicity, $10.4$ cm$^{-1}$, of the bimolecular KRb reaction (Fig.~\ref{fig1}). Further, their translational energy ratio, $0.49\pm 0.06$, is consistent with the expected ratio, 0.46, originating from two different mass products flying apart with zero center-of-mass momentum. This provides further evidence that supports the identification of K$_2^+$ and Rb$_2^+$ ions as arising from ionization of the products of the KRb + KRb chemical reaction.

\begin{figure}
\centering
\includegraphics[width=6 in]{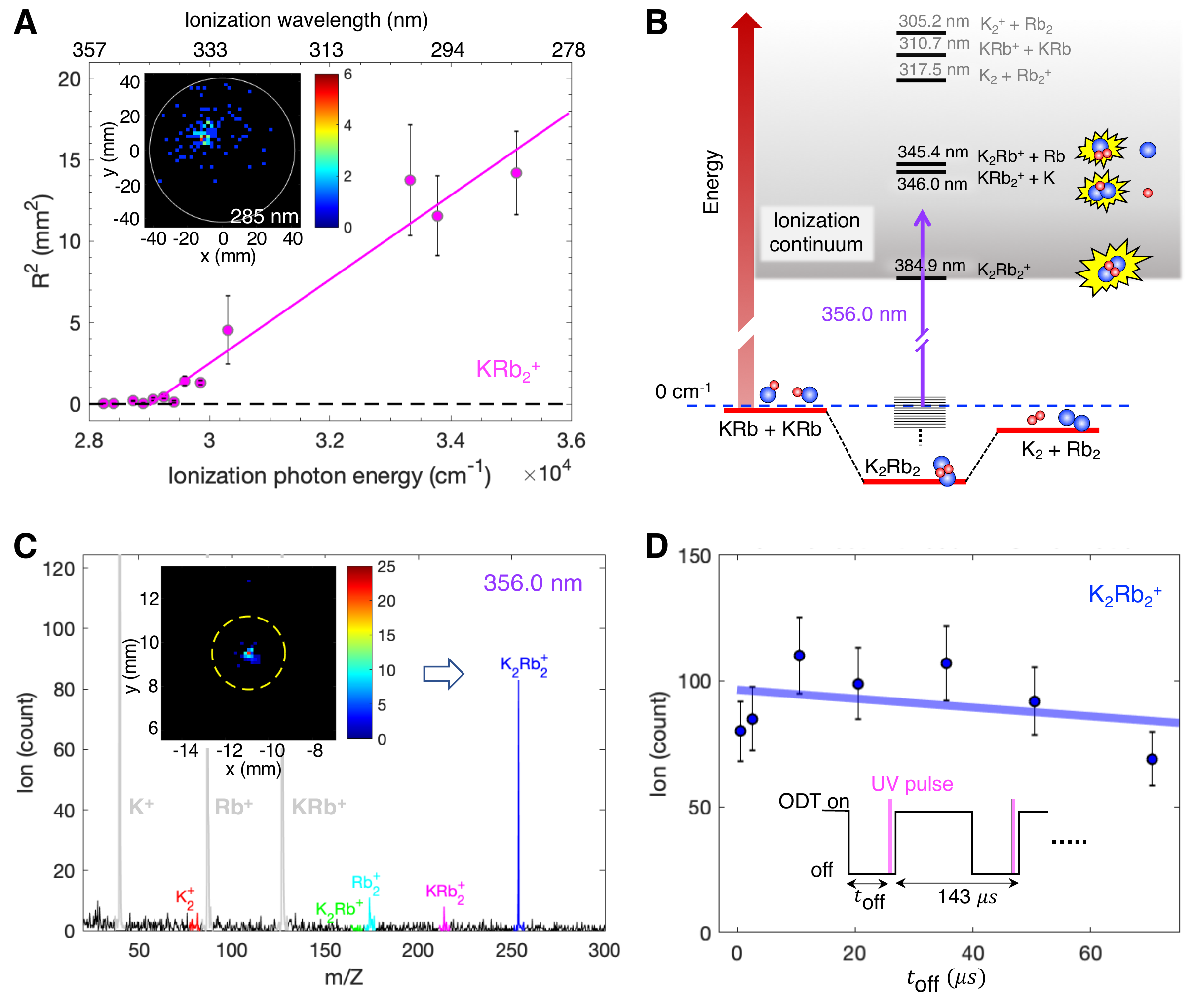}
    \caption{\textbf{Direct detection of the intermediate complex K$_2$Rb$_2^*$.} (\textbf{A}) (Inset) The VM image of detected KRb$_2^+$ ions (using an ionization laser wavelength of 285 nm). For each wavelength, $R^2$ is extracted from such an image, where $R$ is the Gaussian width of the ion spatial distribution, and $R^2$ is proportional to the TER. The measured TER of KRb$_2^+$ ions is plotted versus the ionization photon energy. Error-bars denote the standard deviation of the mean (standard error). Fits are described in S5 (SM). The solid line is an unweighted linear fit to the data above $2.9\times10^4$ cm$^{-1}$ with the RMSE being 1.44, from which an experimental dissociative ionization threshold wavelength of $345\pm 4$ nm is determined. (\textbf{B}) Calculated threshold wavelengths of the direct photoionization and dissociative ionization of the intermediate complex. The energies for the dissociative ionization thresholds are those corresponding to the equilibrium geometry of the ionic complex (table S1) and are therefore lower bounds on the ionization energy. (\textbf{C}) TOF mass spectrum produced using an ionization laser wavelength of 356 nm. (Inset) The corresponding VM image of the detected K$_2$Rb$_2^+$ ions. The yellow dashed circle corresponds to $10.4\text{ cm}^{-1}$.  We do not observe any larger species beyond K$_2$Rb$_2$, up to $m/Z = 1500$. (\textbf{D}) K$_2$Rb$_2^+$ counts are plotted against $t_\text{off}$, where $t_\text{off}$ denotes the length of ODT off time before UV photoionization. Error bars include shot noise and $10\%$ molecule number fluctuations. A weighted linear fit (blue line) with the RMSE being 1.17 determines a slope of $-0.2\pm0.2$, consistent with zero. }
\label{fig4}
\end{figure}

Next, we focused on the transient intermediate collision complex, K$_2$Rb$^*_2$. In order to observe the complexes which by conservation of momentum should only exist in the vicinity of the reactants, we shaped the UV ionization beam into a Gaussian beam profile. After data accumulation, we observed signals consistent with the masses of K$_2$Rb$^+$ and KRb$_2^+$ (fig.~S2). Based on their VM images, which show large translational energies (Fig.~\ref{fig4}A inset), we hypothesize that these ions are  from dissociative ionization of K$_2$Rb$^*_2$. To substantiate this idea, we varied the wavelength of the ionization beam in order to determine the relationship between the translational energy of the triatomic ions and the energy of the photon. We found that the characteristic translational energy associated with the KRb$_2^+$ ion decreases as the ionization energy decreases. The ionization energy where the translational energy becomes zero (at $345\pm 4$ nm) agrees with our theoretical predictions ($346\pm 2$ nm) of the dissociative ionization threshold for the transient intermediate, K$_2$Rb$^*_2$ + h$\nu \rightarrow$ KRb$_2^+$ + K(4s) + $e^{-}$. 

These theoretical calculations of ionization threshold energies of diatomic, triatomic, and tetratomic K- and Rb-containing molecules (shown in Fig.~\ref{fig4}B) are based on the same methodology used in \cite{vexiau2017dynamic} and references therein. Briefly, each alkali-metal atom was modeled as a one-electron system in the field of an ionic core (K$^+$ or Rb$^+$). We used a semi-empirical effective core potential plus a core polarization potential to represent the correlation between the valence electron and the core electrons \cite{SM}. The K$_2$Rb$^+$ and KRb$_2^+$ triatomic ions were modeled as two-valence-electron systems, and the K$_2$Rb$_2^+$ ion as a three-valence-electron molecule. In the framework of such a simplification, the ground-state potential energy surface (PES) can be obtained with good accuracy via the diagonalization of the full electronic Hamiltonian (i.e. full configuration interaction) expressed on a large Gaussian basis set. For all molecular and atomic species, the energies were computed with respect to the same origin, namely the energy of the four cores (K$^+$ + K$^+$ + Rb$^+$ + Rb$^+$). This allowed for the determination of transition energies between different species.

To directly observe the transient intermediate complex K$_2$Rb$^*_2$, we tuned the wavelength of our ionization laser to 356 nm, with energy well below the lowest dissociative ionization channel. Figure~\ref{fig4}C displays a mass spectrum obtained with ionization at 356 nm, and a strong signal of K$_2$Rb$_2^+$ is evident. We emphasize that the ionization process transforms the transient intermediate into a bound molecular ion that has no energetically allowed dissociation channel (Fig. \ref{fig4}B) and can therefore survive its flight to the MCP. Although we have not yet directly measured the lifetime of the complex due to the technical challenges of precisely  establishing a zero of time, the signal strength of our direct observation puts an estimate of a lifetime of 350 ns (or 3~$\mu$s), assuming the ionization cross-section of the K$_2$Rb$_2$ intermediate complex is 10~Mb (or 1~Mb). This cross-section has not been reported in the literature. 

The origin of the observed intermediate complex has been the subject of previous debate~\cite{mayle2013scattering,christianen2019trapping}. The long-lived transient complex could potentially collide with another KRb, causing the prior's decay into a deeply-bound K$_2$Rb$_2$ molecule and leading to the conversion of its internal energy into a large, observable TER~\cite{mayle2013scattering,christianen2019quasiclassical}. In  contrast, we observe a detector resolution-limited small momentum distribution of the K$_2$Rb$_2^+$ ions (Fig.~\ref{fig4}C inset), consistent with the zero-momentum transient intermediate.

Moreover, because the reactants are trapped in the ODT, a light-assisted process could be a competing, confounding factor, as suggested by Christianen \textit{et al.}~\cite{christianen2019trapping}. To examine the role of ODT on the detected intermediate complex, we varied the length of time that the ODT was switched off prior to ionization, from 1 $\mu$s to 70 $\mu$s. If the ODT contributed to the formation of deeply-bound K$_2$Rb$_2$ molecules, which have no radiative decay pathway and only potentially leave the probed volume on a millisecond time scale if they are untrapped, the K$_2$Rb$_2$ would steadily build up in concentration in the presence of the ODT. As a result, the concentration of K$_2$Rb$_2$ should decrease monotonically as we increase the ODT off duration. Instead, we find the yield of K$_2$Rb$_2^+$ ions has no monotonic trend with the ODT off duration (see Fig.~\ref{fig4}D). This result is evidence that the intermediates we observe are formed upon collision of two KRb molecules, with no significant effect from the ODT, on or off.

The direct observation of 2 KRb $\rightarrow$ K$_2$Rb$_2^*$ $\rightarrow$ K$_2$ + Rb$_2$ opens numerous possibilities of exploring the detailed role of quantum mechanics in ultracold chemical reaction dynamics by measuring the lifetime of the intermediate complex~\cite{mayle2013scattering,christianen2019quasiclassical}, testing the transition from quantum to semiclassical reactions~\cite{gao2010universal}, and resolving the quantum states of the reaction products~\cite{croft2017universality} and the intermediate.

\textbf{Acknowledgments:} We thank D. Herschbach, L. Zhu, T. Karman, and J. Ye for discussion, K. Liu for introducing us to the VMI techniques, T. Pfau, E. Narevicius and M. Greiner for  discussions on apparatus design,  J. Doyle for loaning   laser equipment, and  W. Stwalley, P. Gould and the late E. Eyler for sharing KRb spectroscopy literature. The $^{40}$K isotope used in this research was supplied by the United States Department of Energy Office of Science by the Isotope Program in the Office of Nuclear Physics. \textbf{Funding:} This work is supported by  the DOE Young Investigator Program, the David and Lucile Packard Foundation, and the NSF through Harvard-MIT CUA. \textbf{Author contributions:} The experimental work and data analysis were carried out by M.-G.H., Y.L., D.D.G., Y.-W.L., A.H.G., T.R., and K.-K.N.. Theoretical calculations were done by R.V., N.B., and O.D.. All authors contributed to interpreting the results and writing the manuscript. \textbf{Competing interests:} The authors declare that they have no competing financial interests. \textbf{Data and materials availability:} Data from the main text and supplementary materials are available through the Harvard Dataverse at~\cite{DVN/MDFLOZ_2019}.

\bibliography{refs}

\begin{thebibliography}{41}
\bibitem [{foo()}]{footnote1}%
  \BibitemOpen
  \href@noop {} {\bibinfo  {journal} {170 $\mu$s is the time it would take for
  a $\textrm{Rb}_2$ molecule with translational energy of 0.0127
  $\text{cm}^{-1}$ to travel from the center of the KRb cloud to the ionization
  ring with a diameter of 0.45 mm}\ }\BibitemShut {NoStop}%
\end{thebibliography}
\end{document}


\author{M.-G.~Hu}
\thanks{These two authors contributed equally.}
\affiliation{Department of Chemistry and Chemical Biology, Harvard University, Cambridge, Massachusetts, 02138, USA.}
\affiliation{Department of Physics, Harvard University, Cambridge, Massachusetts, 02138, USA.}
\affiliation{Harvard-MIT Center for Ultracold Atoms, Cambridge, Massachusetts, 02138, USA.}

\author{Y.~Liu}
\thanks{These two authors contributed equally.}
\affiliation{Department of Physics, Harvard University, Cambridge, Massachusetts, 02138, USA.}
\affiliation{Department of Chemistry and Chemical Biology, Harvard University, Cambridge, Massachusetts, 02138, USA.}
\affiliation{Harvard-MIT Center for Ultracold Atoms, Cambridge, Massachusetts, 02138, USA.}

\author{D.~D.~Grimes}
\affiliation{Department of Chemistry and Chemical Biology, Harvard University, Cambridge, Massachusetts, 02138, USA.}
\affiliation{Department of Physics, Harvard University, Cambridge, Massachusetts, 02138, USA.}
\affiliation{Harvard-MIT Center for Ultracold Atoms, Cambridge, Massachusetts, 02138, USA.}

\author{Y.-W.~Lin}
\affiliation{Department of Chemistry and Chemical Biology, Harvard University, Cambridge, Massachusetts, 02138, USA.}
\affiliation{Department of Physics, Harvard University, Cambridge, Massachusetts, 02138, USA.}
\affiliation{Harvard-MIT Center for Ultracold Atoms, Cambridge, Massachusetts, 02138, USA.}

\author{A.~H.~Gheorghe}
\affiliation{Department of Physics, Harvard University, Cambridge, Massachusetts, 02138, USA.}

\author{R. Vexiau}
\affiliation{Laboratoire Aim\'{e} Cotton, CNRS, Universit\'{e} Paris-Sud, ENS Paris-Saclay, Universit\'{e} Paris-Saclay, 9145 Orsay cedex, France}

\author{N. Bouloufa-Maafa}
\affiliation{Laboratoire Aim\'{e} Cotton, CNRS, Universit\'{e} Paris-Sud, ENS Paris-Saclay, Universit\'{e} Paris-Saclay, 9145 Orsay cedex, France}

\author{O. Dulieu}
\affiliation{Laboratoire Aim\'{e} Cotton, CNRS, Universit\'{e} Paris-Sud, ENS Paris-Saclay, Universit\'{e} Paris-Saclay, 9145 Orsay cedex, France}

\author{T.~Rosenband}
\affiliation{Department of Physics, Harvard University, Cambridge, Massachusetts, 02138, USA.}

\author{K.-K.~Ni}
\email[To whom correspondence should be addressed. E-mail: ]{ni@chemistry.harvard.edu}
\affiliation{Department of Chemistry and Chemical Biology, Harvard University, Cambridge, Massachusetts, 02138, USA.}
\affiliation{Department of Physics, Harvard University, Cambridge, Massachusetts, 02138, USA.}
\affiliation{Harvard-MIT Center for Ultracold Atoms, Cambridge, Massachusetts, 02138, USA.}

\title{{\Large Supplementary Materials: Direct observation of bimolecular reactions of ultracold KRb molecules}}
\maketitle
\large

\section{Photoionization and velocity-map imaging of ions}
\begin{figure*}
\centering
\includegraphics[width=6.0 in]{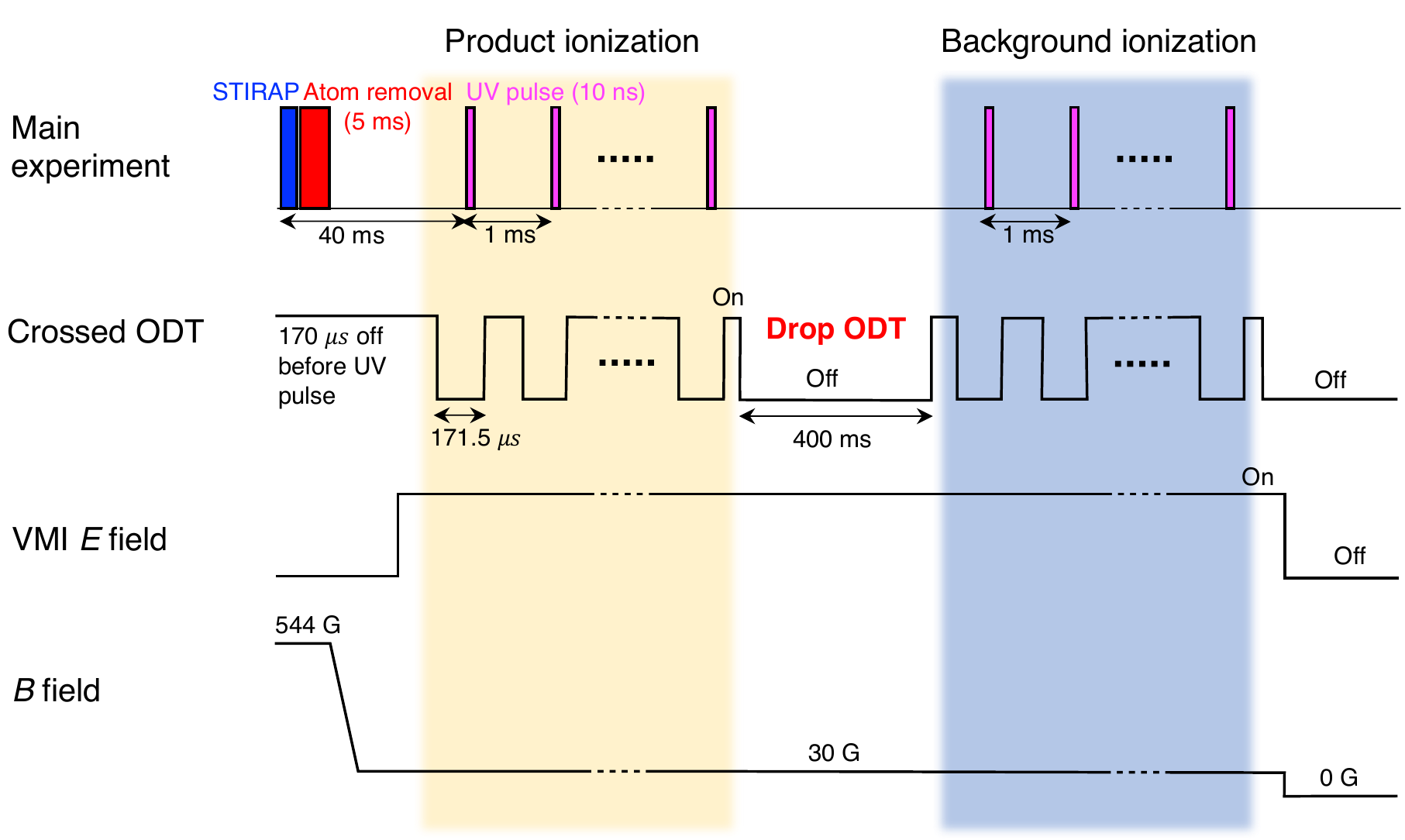}
    \caption{\textbf{Timing diagram for photoionization detection of an ultracold chemical reaction.} When VMI electrodes are on, it generates a 180 V/cm electric field at the location of the ODT. The ODT is turned off for 400 ms between the product ionization stage and the background ionization stage to ensure that all particles inside the ODT drop out under gravity.}
\label{fig_timing}
\end{figure*}

\begin{figure*}
\centering
\includegraphics[width=4 in]{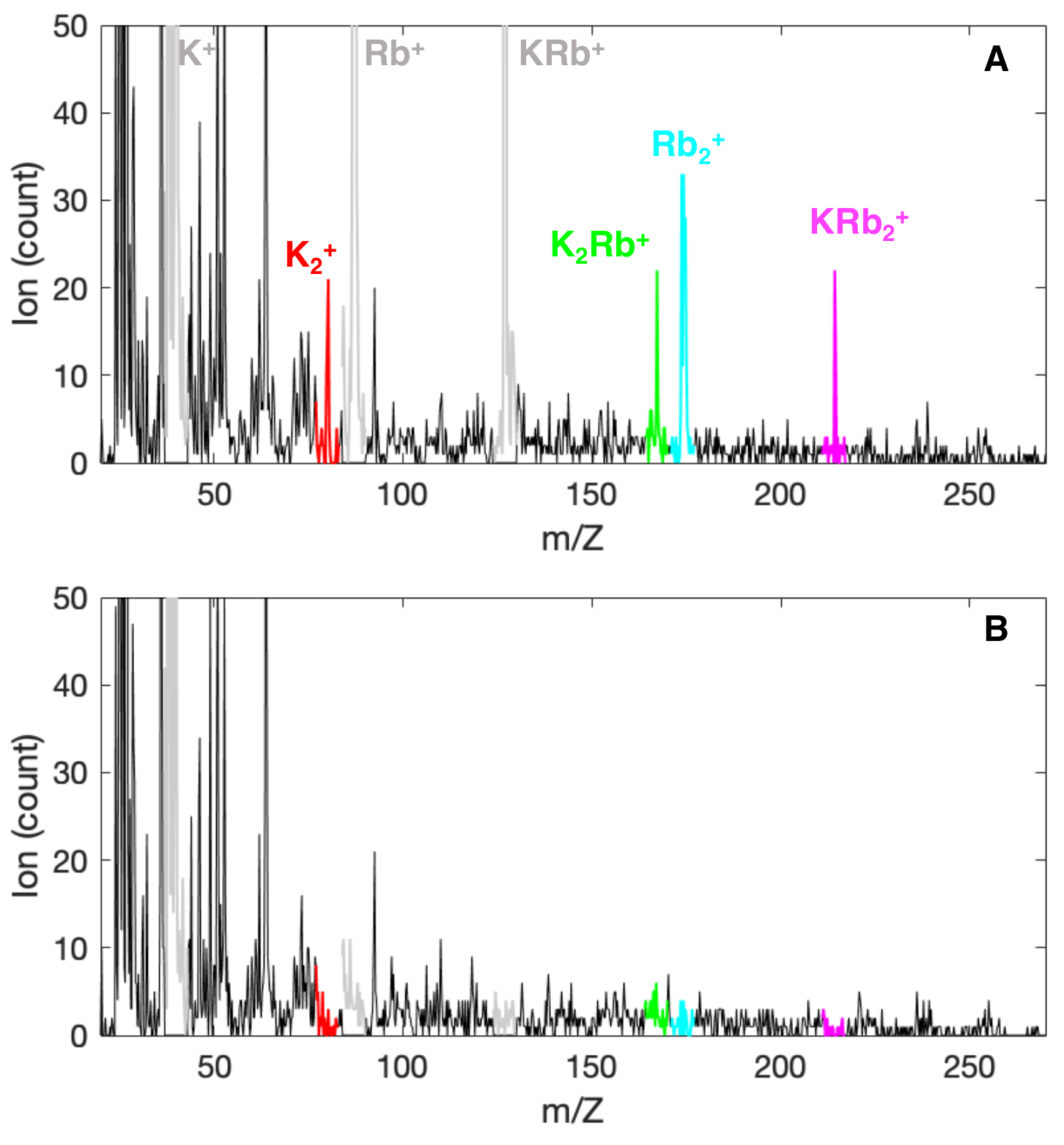}
    \caption{\textbf{Mass spectra recorded as the reaction is probed by a 285 nm UV laser with a Gaussian beam profile.} (\textbf{A}) Signal spectrum. (\textbf{B}) Background spectrum.}
\label{figS2}
\end{figure*}

Our photoionization laser system is a frequency-doubled, broadly tunable pulsed dye laser (LIOP-TEC/LiopStar-HQ) pumped by a pulsed Nd:YAG laser (EdgeWave BX80). It has a variable repetition rate of up to 10 kHz, a pulse duration of 7 ns, a spectral width of 0.06 cm$^{-1}$, and a tuning range of 220 - 400 nm (after frequency-doubling by a BBO3 crystal). Once the ground state KRb molecules are created in the ODT, they begin to react with each other. In order to perform photoionization detection on the reaction products, we follow the sequence of events shown in fig.~\ref{fig_timing}. The sequence for detecting the intermediate is the same except for the repetition rate and ODT switch off time (see inset of Fig.~4D). To distinguish signal from background noise, we take two TOF mass spectra in succession for all of the experiments carried out in this study. The ODT is switched off for 400 ms between the recordings of the two spectra to ensure that all particles in the ODT are dropped out before we take the background spectrum. An example set of spectra is shown in fig.~\ref{figS2}, where ion signals are recorded as the reaction is probed by a 285 nm UV laser with a Gaussian beam profile. By comparing the spectra in fig.~\ref{figS2}A and B, we can clearly distinguish the signal mass peaks from the background.

We designed and built VMI ion optics in order to project the momentum/translational energy distribution of the photoionized neutrals into a spatial distribution that can be detected by a position sensitive delay-line MCP (RoentDek/DLD80). The MCP has an active detection area of $ \varnothing~80 $ mm, a spatial resolution of 0.08 mm, and a temporal resolution of 20 ns. A VMI setup consists of three main electrode plates: a repeller plate, an extractor plate, and a ground plate, with the geometries and voltages of these electrodes  chosen to form the VMI configuration \cite{eppink1997velocity}. Before ionization, a small volume of neutral reaction products sits between the repeller and the extractor. After ionization, charged species are accelerated towards the ion detector along the TOF axis while they ballistically expand away from that axis at a rate determined by their transverse velocities. The spatial distribution of ions on the detector associated with each species forms its VM image. The transverse translational energy ($KE$) of each ion is mapped to its radial distance $R$ from the center of the image according to the relation $R = A\cdot\sqrt{KE/V_R}$, where $V_R$ is the repeller voltage and $A = 15.88$ mm$/\sqrt{\text{cm}^{-1}/\text{V}}$.

\begin{figure*}
\centering
\includegraphics[width=5.0 in]{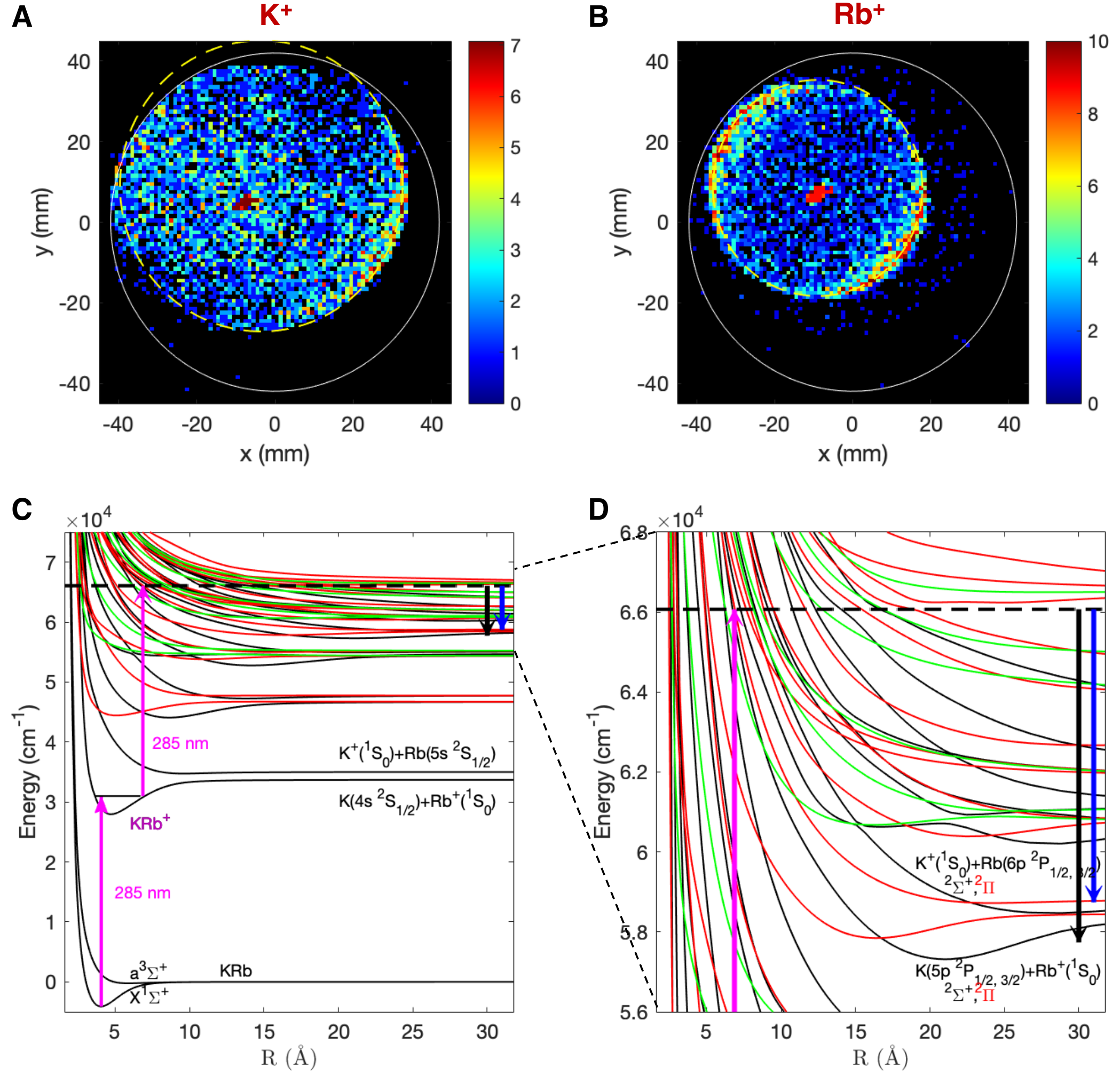}
    \caption{\textbf{Photoionization of GS KRb molecules.} (\textbf{A-B}) Measured VMI signals of K$^+$ and Rb$^+$ ions at 285 nm and $V_R = 1000$ V. The white circles represent the detector active area. The yellow dashed circles correspond to the momenta of the K$^+$ and Rb$^+$ ions acquired during the dissociative ionization of ground state KRb molecules. The radius of the yellow circle for the K$^+$ (Rb$^+$) ions is 37 mm (26 mm) corresponding to TER of 0.73$\times 10^4$ cm$^{-1}$ (0.83$\times 10^4$ cm$^{-1}$). (\textbf{C-D}) Energy diagram for KRb and KRb$^+$ molecules. The potential energies of the electronic states of molecular KRb$^+$ ions are taken from~\cite{korek2003theoretical}. The magenta arrows represent the photon energy of the ionization laser. KRb$^+$ ions are generated via single-photon ionization. In addition, K$^+$ and Rb$^+$ ions are generated via two-photon dissociative ionization, during which the released energy is distributed between Rb$^+$ (Rb) and K (K$^+$) according to their mass ratio. The measured TER, 0.73$\times 10^4$ cm$^{-1}$, from the VMI signal of K$^+$ ions in (\textbf{A}) is drawn as the blue arrow above, which matches the energy of the dissociative ionization channels, K$^+$($^1$S$_0$) + Rb(6p~$^2$P$_{1/2,3/2}$). The measured TER, 0.83$\times 10^4$ cm$^{-1}$, from the VMI signal of Rb$^+$ ions in (\textbf{B}) is drawn as the black arrow above, which matches the energy of the dissociative ionization channels, K(5p~$^2$P$_{1/2,3/2}$) + Rb$^+$($^1$S$_0$).}
\label{figS3}
\end{figure*}

For chemical reactions that occur at less than 1~$\mu$K, the translational energy of the reactants is negligibly small. Therefore, the momentum of the ions enables us to determine if the ions are generated via direct or dissociative photoionization. The former only detaches an electron, and thus causes no measurable  change on translational energy of ions. The latter involves breaking chemical bonds and generally involves a significant increase of ion translational energy. As shown in fig.~\ref{figS3}A-B, from the VMI signals of both K$^+$ and Rb$^+$ ions, we observe two features: an isotropic central peak and an anisotropic ring. Because the ions from the central peak have a very small translational energy, we identify them as produced via single-photon ionization of the ultracold K and Rb atoms left in the ODT. This identification is further confirmed by the observation that these small features disappear when we tune the UV photon energy below the atomic ionization thresholds of K and Rb. For ions that form the ring, we can determine the translational energy release (TER) from the radius of the ring, and the energy matches well to the dissociative ionization potentials of KRb (see fig.~\ref{figS3}C-D). Therefore, we identify these ions as  originating from dissociative ionization of the ultracold KRb molecules.

\begin{table*}
\caption{\label{tab:1PI} \textbf{Ionization information for all relevant species in the experiment.} We define the photoionization (PI) threshold here as the energy required to excite from the ground ro-vibronic state of the neutral to the ground ro-vibronic state of the ion. The photoionization cross section is shown for excitation at 285 nm.}
\begin{tabular}{lccc}
\hline\hline
Species & PI threshold (nm) & PI cross section $\sigma$ (Mb$^*$) & MCP detector efficiency$^\dag$\\
\hline
K & $285.6$ (Ref.~\cite{hudson1965absorption}) & 0.01 (Ref.~\cite{hudson1965absorption}) & 0.394\\
Rb & $296.8$ (Ref.~\cite{suemitsu1983relative})  &0.049 (Ref.~\cite{suemitsu1983relative}) & 0.355\\
K$_2$ & $305.1^\ddag$ (Ref.~\cite{broyer1983k2})  & $0.50 \pm 0.25$ (Ref.~\cite{creek1968some}) & 0.360 \\
Rb$_2$ & $317^\ddag$ (Ref.~\cite{suemitsu1983relative}) & $0.58 \pm 0.26$ (Ref.~\cite{creek1968some,suemitsu1983relative}) & 0.310\\
KRb & $310.0-320.3^\ddag$ (Ref.~\cite{kappes1985generation})  & & 0.331\\
K$_2$Rb & $374.9^\S$ (this work)& & 0.313 \\
KRb$_2$ & $377.6^\S$  (this work)&  & 0.295\\
K$_2$Rb$_2$ & $347.65^\P$ (this work)&  & 0.282\\
\hline\hline
\end{tabular}
$^*$~1 Mb = $10^{-18}$ cm$^2$.\\
$^\dag$~Values are calculated using the empirical formula developed in Ref.~\cite{krems2005channel} based on the ion's mass and translational energy. We take into account both the intrinsic efficiency of the channel plates (60$\%$ open-area-ratio) of our MCP and the 75$\%$ optical transmission of a mesh in front of the plates. The maximum detection efficiency, in the limit of light mass and high translational energy, is $0.60 \times 0.75 = 0.45$.\\
$^\ddag$~It is calculated to be $305.2$ nm for K$_2$, $317.5$ nm for Rb$_2$, and 310.72 nm for KRb in this work.  \\
$^\S$~These values are calculated for the equilibrium geometries of the ionic complexes
KRb$_2^+$ ($C_{2v}$): $R=7.15$ a.u., $r_{\text{Rb}_2}=8.88$ a.u., $\theta=90^\circ$ and RbK$_2^+$ ($C_{2v}$): $R=7.51$ a.u., $r_{\text{K}_2}=8.08$ a.u., $\theta=90^\circ$. Here we use Jacobi coordinates: $r$ is distance of the homonuclear pair and $R$ is distance from the third partner to the center of the homonuclear pair. The threshold for $\text{K}_2\text{Rb}_2^*+h\nu\rightarrow \text{K}_2\text{Rb}^++\text{Rb}(5S)+e^-$ is 345.4 nm and the threshold for $\text{K}_2\text{Rb}_2^*+h\nu\rightarrow \text{K}\text{Rb}_2^++\text{K}(4S)+e^-$ is 346.0 nm.  \\
$^\P$~This value is calculated for the equilibrium geometry of the ionic complexes
K$_2$Rb$_2^+$ (T-shaped): $R=11.8$ a.u., $r_{\text{K}_2}=r_{\text{Rb}_2}=8.5$ a.u.. Here $R$ is distance between K$_2$ center and Rb$_2$ center. The threshold for K$_2$Rb$_2^*+h\nu\rightarrow$ K$_2$Rb$_2^++e^-$ is 384.9 nm.    
\end{table*}

In table \ref{tab:1PI}, we summarize the ionization energy thresholds of relevant molecular species to this work. Experimentally measured thresholds are available for the atomic and diatomic species from literature. The triatomic and tetratomic thresholds are derived from the differences between the energies of their respective ion and neutral ro-vibronic ground states. The neutral triatomic and tetratomic energies are calculated in~\cite{byrd2010structure}. The ionic triatomic and tetratomic energies are calculated in this work using the following methods. The molecular species are described in Jacobi coordinates. We did not derive here the full PES of their electronic ground state. Instead, inspired by previous results on metallic clusters~\cite{ray1989ab,ray1990ab}, we first looked at each potential local minimum of the molecular species: $C_{2v}$ and linear for the triatomic ion, diamond, T-shaped and linear for the tetratomic ion, with each possible arrangement of the individual atoms. We computed the potential energy on a sparse grid (steps of 0.3 a.u. in each radial distance, the angles being fixed by the selected geometry), and compared the values for the different geometries. Once the global minimum is found a more accurate calculation is made (steps of 0.02 a.u.) to pinpoint the potential energy at the equilibrium geometry.

\section{Ionization beam geometry}
The output of the pulsed UV laser is shaped into a $\varnothing~3$ mm beam. To detect the reactants or intermediate complex we shrink down the beam to create a $\varnothing~300$ $\mu$m Gaussian profile at the molecule location. To detect the products we generate hollow-bottle beams following the optical system design from~\cite{Kulin_2001}. Each hollow beam has a $\varnothing~0.45$ mm ring-shaped profile centered around the KRb cloud, with a Gaussian width of $5.9~\mu$m. We generate two such hollow beams and cross them at a $40^{\circ}$ angle centered on the ODT to maximize the capture solid angle of the reaction products. Due to a combination of limited numerical aperture, finite sharpness of the tip of the axicon, and the spatial mode of the input beam, the center of each hollow beam is not perfectly dark. We estimate the intensity at the center of the hollow beam to be $\sim 0.2\%$ of its peak intensity. This explains the presence of in-trap species (K, Rb, and KRb) in our mass spectrum (Fig.~3A) acquired with the crossed hollow beam.

\section{Modeling the time-of-flight lineshapes of reaction products}

Species ionized in the ring of peak intensity of the hollow beam have a different TOF lineshape
than species ionized in the weak center of the hollow beam. Ions that are generated closer to the MCP arrive later and those generated further from the MCP arrive earlier. This is because ions that are generated closer to the MCP experience less time in the accelerating electric fields and so have a lower velocity in the field-free drift region. In order to simulate the TOF distribution for ions generated in the ring, we performed ion trajectory simulation in SIMION~\cite{eppink1997velocity} with the realistic electric field created by considering the voltages and geometries of the VMI ion optics. For each product species, 10,000 ions are sampled according to the  density distribution of the product as well as the ionization laser fluence distribution. Each simulated TOF distribution is convolved with a Gaussian function (sigma = 4.2 ns) reflecting the time resolution of our MCP. The resulting TOF lineshape is then fitted to the corresponding experimental data, with the overall amplitude as the only fitting parameter. The results are shown in Fig. 3D and E. The agreement between the widths and overall shapes of the experimental and simulated distributions supports our understanding for the spatial origins of the product ions. We speculate that the presence of the center peak, prominently seen in our data in figure 3E yet not captured by the simulated lineshape, is due to the non-zero intensity at the center of the hollow beams ionizing high density products at and around the center of the reactant KRb cloud.  This center peak is less evident for K$_2^+$ because it has fewer ion counts and a narrower feature width, both of which make the center peak harder to resolve.  The imperfections in the hollow beam are difficult to model exactly and are thus not included in the simulation.

\section{Ion-neutral collisions}

The rate of ion-neutral collisions can be estimated by using a long-range capture model to determine the Langevin rate coefficient. The rate coefficient for ion-dipole collisions is~\cite{clary1990fast}:
\begin{equation}
    k_{\text{Ion-Dipole}}(T) = \frac{\mu_Dq}{4}\left[\frac{8\pi}{\mu}\right]^{\frac{1}{2}} \left(k_BT\right)^{-\frac{1}{2}} \frac{1}{4\pi\epsilon_0}
\end{equation}
\noindent where $\mu_D$ is the dipole moment of the molecule, $q$ is the charge of the ion, $\mu$ is the reduced mass of the ion-dipole system, and $T$ is the energy of the collision in units of Kelvin. A KRb$^+$ ion generated in the center of the ODT has exited the trap (using the 4$\sigma$ width as the trap dimension) within 70 ns of being created. As the KRb$^+$ ions are accelerated towards the MCP, their Langevin rate coefficients rapidly decrease as the translational energy of the ion increases. The average number of collisions that a KRb$^+$ ion formed in the center of an ODT with a density of neutral KRb molecules of $10^{12}$ cm$^{-3}$ experiences is $2.5\times10^{-5}$. 

Ion-neutral collisions could be a potentially confounding factor in all of the experiments we describe in the main text. A collision between KRb and KRb$^+$ is potentially reactive and can lead to the formation of diatomic, triatomic, or tetratomic ions of the same species as the ones we observe in the experiments. To place an upper bound on the number of ion-neutral reaction products generated in each experimental cycle, we estimate the number of KRb$^+$ produced by the photoionization of KRb, calculate the number of ion-neutral collisions, and make the assumption that every such collision leads to the formation of a product ion. These estimates are compared to our measured number of ions per experimental cycle below.

Varying experimental conditions (repetition rate, ionization beam geometry, ionization laser power, ionization wavelength) between different experiments leads to differing amounts of KRb$^+$ generated and therefore differing rates of ion-neutral collisions. The K$_2$ and Rb$_2$ product experiment (Fig.~3) has a maximum creation rate of product molecular ions via ion-neutral collisions of $7.2\times10^{-3}$ per experimental cycle, compared to the observed values of 0.82 K$_2^+$ and 4.2 Rb$_2^+$ per experimental cycle. The dissociative ionization trimer experiment (Fig.~4A) has a maximum creation rate of trimer molecular ions via ion-neutral collisions of $1.4\times10^{-3}$ per experimental cycle, compared to the observed values of 0.03 per experimental cycle for both KRb$_2^+$ and K$_2$Rb$^+$. The transient intermediate complex experiment (Fig.~4C) has a maximum creation rate of tetramer molecular ions via ion-neutral collisions of $1.0\times10^{-4}$ per experimental cycle, compared to the observed value of 1.2 per experimental cycle. Based on these comparisons, we conclude that ion-neutral reactions do not significantly contribute to the measured ion signals in any of our experiments.

\section{Bayesian analysis of VM image}
Bayesian fits \cite{sivia2006data} determine the signal radius in each velocity-map image for a model where a circular Gaussian (signal) is added to a constant background (noise). This model, although heuristic, appears to fit the data well.

Each fit has five parameters (background amplitude $b$; Gaussian amplitude $A$, horizontal center $x_0$, vertical center $y_0$, radius $R$), collectively labeled $\pmb{\theta}=(b,A,x_0,y_0,R) \in \mathbb{R}^5$. We calculate the probability of occurrence for a measured pixelated image pair (signal image and background image) as the product of individual pixel probabilities
\begin{equation*}
P_{images}(\pmb{\theta}) = \prod_{\substack{pixels\\ t \in {0,1}} } P_{pixel}(x,y,t,n(x,y),\pmb{\theta})
\end{equation*}
where $n(x,y)$ is the non-negative integer number of counts at location $x,y$ in the measured image, $t=0$ for signal and $1$ for background and 
\begin{equation*}
P_{pixel}(x,y,t,n,\pmb{\theta}) = \frac{\lambda(x,y,t,\pmb{\theta})^n}{n!}e^{-\lambda(x,y,t,\pmb{\theta})}
\end{equation*}
is the Poisson probability of each individual pixel accumulating the measured counts $n$. The rate parameter $\lambda$
for location $x,y$ at time $t$ and model parameters $\pmb{\theta}$ is
\begin{equation*}
\lambda(x,y,t,\pmb{\theta}) = 
\begin{cases}
b + \frac{A}{2\pi R^2} e^{-\frac{(x-x_0)^2+(y-y_0)^2}{2R^2}}, & \text{if } t = 0,\\
b, & \text{if } t = 1.
\end{cases}
\end{equation*}
At the beginning of the measurement ($t = 0$), the Poisson rate $\lambda$ contains signal and background, while at the end of the measurement ($t = 1$), the signal source has been depleted by many repetitions of photoionization and only background remains.

\begin{table*}
\caption{\label{tab:Bayesian} \textbf{Number of KRb$_2^+$ ions used in the Bayesian analysis for Fig. 4A.} }
\begin{center}
\begin{tabular}{ccc}
\hline\hline
Ionization laser & Ion number& Ion number  \\
wavelength (nm)&from signal image&from background image\\
\hline
285 & 79   & 17\\
296 & 44   & 15\\
300 & 25   & 7  \\
330 & 8     & 2  \\
335 & 281 & 78\\
338 & 102 & 18\\
340 & 17   & 3\\
342 & 34   & 9\\
344 & 21   & 5\\
346 & 18   & 4\\
348 & 54   & 9\\
352 & 18   & 5\\
354 & 55   & 17\\
\hline\hline
\end{tabular} 
\end{center}
\end{table*}

Based on Bayes' theorem, the probabilities $P_{images}(\pmb{\theta})$ together with uninformative priors ($log(b), log(A), log(R), x_0, \text{and } y_0$ are each uniformly distributed over all $\mathbb{R}$) for the fit parameters $\pmb{\theta}$ yield the 5-D probability density $p(\pmb{\theta})$.  Numerical integration (marginalization) of this function over the four other fit parameters leaves the probability density for the radius $R$, from which the mean and the standard deviation of the mean (standard error) is calculated~\cite{sivia2006data}. Figure 4A in the main text shows these values measured at different wavelengths and the ion number used for the Bayesian analysis is listed in table~\ref{tab:Bayesian}.

\section{Ion background in the TOF mass spectrum}

\begin{figure*}
\centering
\includegraphics[width=5.0 in]{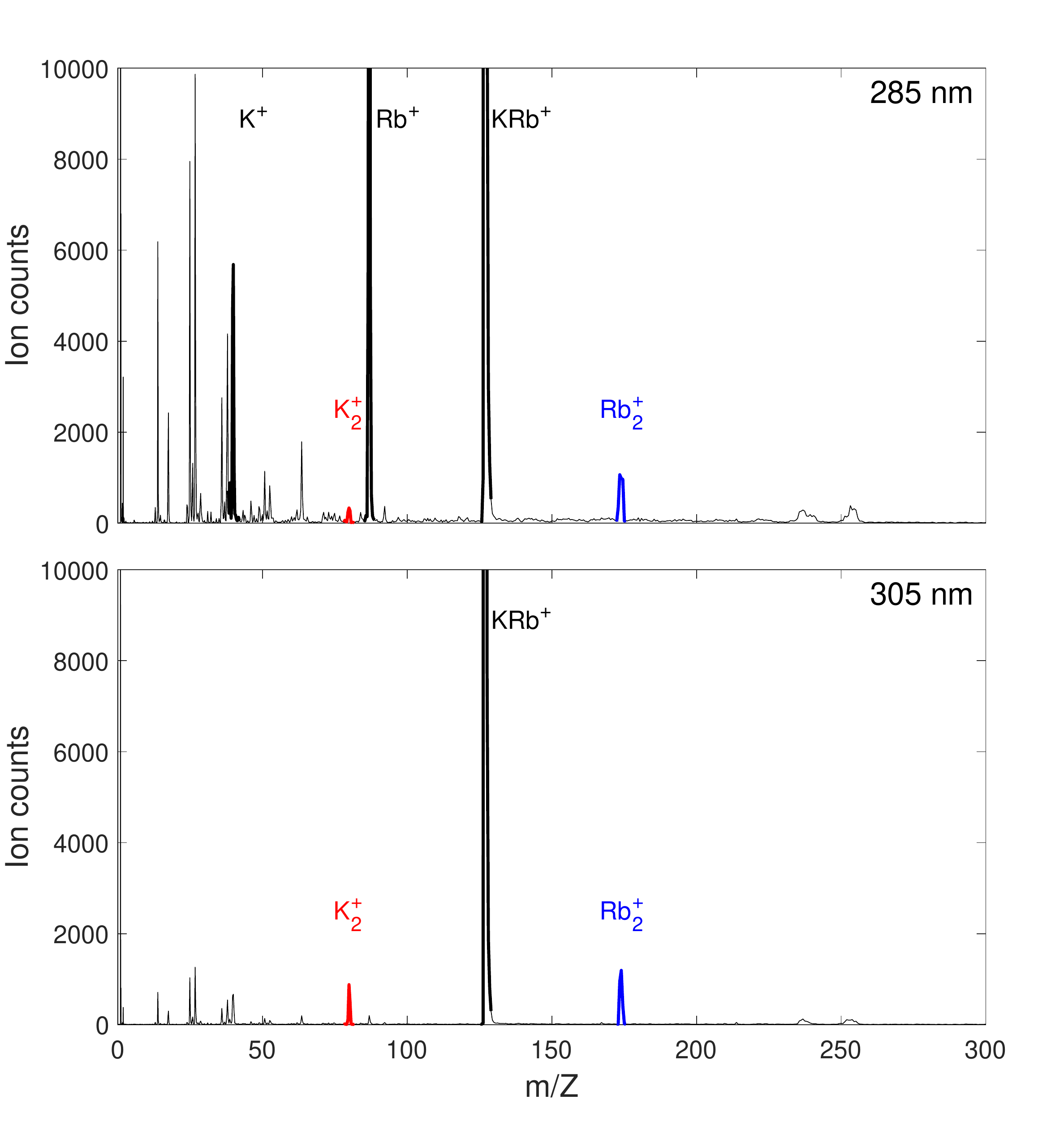}
    \caption{\textbf{Normalized TOF mass spectra acquired using 285 and 305 nm UV laser.} Ion counts in each spectrum is normalized by the pulse energy, repetition rate, and duration of the UV laser, as well as the number of experimental cycles over which the spectrum is accumulated. The masses corresponding to the detected species of interest are color-coded and labeled. The amplitudes of the noise peaks in the 305 nm spectrum are, on average, reduced by a factor of 8 compared to those in the 285 nm spectrum. The K$^+$ and Rb$^+$ peaks are missing in the 305 nm spectrum since the photon energy is below the ionization thresholds of those species.}
\label{figS4}
\end{figure*}

Figure \ref{figS4} shows the normalized mass spectra acquired at UV wavelengths of 285 and 305 nm, respectively, under otherwise similar experimental conditions. The 285 nm spectrum is the same as Fig. 3A save for a re-scaling of the y-axis to better show the low-mass noise peaks. The same set of noise peaks can be identified in both spectra, but their amplitudes, on average, are different by a factor of $8$. As we vary the wavelength of UV ionization laser, we observe a general trend that the background level drops as the wavelength is increased, consistent with photoelectric behavior. We therefore speculate that these noise peaks are due to ions generated by photoelectrons, liberated by scattered UV light and accelerated by the VMI electric field, impacting on objects inside our vacuum chamber (e.g. chamber walls, electric field plates, supporting structures, etc.). Both 285 and 305 nm are above the ionization thresholds of the reaction products. Despite the significantly different levels of background ion counts in the two cases, the measured number of reaction products are roughly consistent with their wavelength-dependent ionization cross sections \cite{suemitsu1983relative, creek1968some}. The VM images of products also show similar distributions for the two cases. We therefore conclude that the background ions present in our experiment does not confound the detection of the species of interest.

\newpage

\bibliography{refs}